# Ferroelectrics and Photovoltaics in Endohedral Fullerenes-based van der Waals Heterostructures

Jie Li and Ruqian Wu*

*Department of Physics and Astronomy, University of California, Irvine, California 92697-4575, USA.*

Using the density functional theory calculations, we studied the ferroelectric, electrical and optical properties of series of bias-controllable endohedral fullerenes ($TM@C_{28}$, TM= Ti-Ni). The important finding is that $Cr@C_{28}$ has a large electric dipole moment and an appropriate bandgap (~1.2eV) as building blocks for ferroelectric-photovoltaic materials. By sandwiching $Cr@C_{28}$ molecules between two van der Waals layers, e.g., graphene or transition metal dichalcogenides, the heterostructures may have exotic ferroelectric and photovoltaic properties with high tunability and performance.

* E-mail: wur@uci.edu

As energy demand continuously increases and global fossil fuels are gradually exhausted, solar energy is an obvious sustainable resource of energy in the foreseeable future. Tremendous research efforts have been dedicated to find ways for effective harvest and utilization of solar energy and diverse approaches have been proposed and designed.[1-5] Photovoltaics that uses semiconductors or their heterostructures to convert solar energy into electricity is one of the most convenient ways.[2,5-7] Many photovoltaic materials have been explored and applied, from crystalline silicon,[8-9] amorphous silicon,[10-11] to more recent perovskites[12-14] and quantum dot solids[15]. The power conversion efficiency is steadily increasing, even though some of these materials are too expensive for large scale deployments.[2,16] New designs for photovoltaic materials are always needed to propel this exciting field forward, and the interest in using novel two-dimensional materials for these designs is rapidly surging.

Meanwhile, ferroelectrics have been identified as promising photovoltaic materials [17-21] as the spontaneous electric polarization promotes the separation of electrons and holes generated by photons. Comparing to the open circuit voltage ($V_{oc}$) in traditional p-n junction solar cells that is limited by the bandgap, the output $V_{oc}$ of a ferroelectric-photovoltaic (FE-PV) material can be much larger than its bandgap.[18] To date, the anomalous photovoltaic effect has been extensively explored in the $LiNbO_3$ family, $PbTiO_3$ family and $BiFeO_3$ family.[19-21] In addition, ferroelectric materials are also stable in a wide range of mechanical, chemical and thermal conditions and can be fabricated using low-cost methods such as sol-gel thin-film deposition and sputtering.[18,22-23] Thus, the FE-PV materials appear to be very promising for the next generation photovoltaic technologies.

As previous works,[24-29] endohedral fullerenes with metal atoms, ions, or clusters embedded in carbon cages can give rise to exotic properties by adjusting their ingredients, size and symmetry, especially for single-molecule ferroelectricity and controllable electronic structures. Meanwhile, graphene and transition metal dichalcogenides (TMDCs) as novel 2D material have astonishing electrical and optical properties, and outstanding scalability and modularity, which makes them play a key role in the fabricating of ultralight solar cells.[30-32] In these solar cells, they can serve multiple functions, such as window electrode, hole/electron collector and extraction, and antireflection layer. Therefore, a heterostructure based on endohedral fullerenes and graphene or TMDCs may offer a potential candidate of ultralight solar cells. More importantly, the ongoing effort of the synthesis of vertical heterostructures based on fullerenes and graphene by self-assembly process has made a rapid progress in experiment,[33-35] and hence there is no big technological barrier to further develop them as long as attractive properties are predicted.

In this work, we systematic investigate the ferroelectric, electrical and optical properties of series of bias-controllable endohedral fullerenes (TM@$C_{28}$, TM= Ti-Ni) by using density functional calculations. The calculations show that Cr@$C_{28}$ has large electric dipole moment and idea bandgap (~1.2eV) as building blocks for ferroelectric-photovoltaic materials. Three van der Waals Heterostructures consist of Cr@$C_{28}$, graphene or transition metal dichalcogenides were furtherly designed, which have astonishing ferroelectric and photovoltaic properties. These findings suggest a new strategy for the designs of FE-PV materials.

All the density functional theory calculations in this work are carried out with the Vienna ab-initio simulation package (VASP) at the level of the spin-polarized generalized-gradient approximation (GGA) with the functional developed by Perdew-Burke-Ernzerhof.[36] The interaction between valence electrons and ionic cores is considered within the framework of the projector augmented wave (PAW) method.[37-38] The Hubbard U of 2.0 eV was adopted to describe the electron correlation in the d-shells of TM cores.[39] The vdW correction (DFT-D3) was included for the description of dispersion forces.[40] The energy cutoff for the plane wave basis expansion is set to 600eV. All atoms are fully relaxed using the conjugated gradient method for the energy minimization until the force on each atom becomes smaller than 0.01 eV/Å, and the criterion for total energy convergence is set at $10^{-5}$ eV.

One possible FE-PV heterostructure is shown in Figure 1(a), which contain of ferroelectric fullerene molecules and two layers of two-dimension (2D) material. The ferroelectric fullerene molecules offer the spontaneous electric polarization and absorb light as well, while the 2D overlayers mostly serve as the hole-transport layer (HTL) and the electron-transport layer (ETL). The maximum internal electric field induced by the intrinsic electric dipole **P** in the heterostructure can be estimated by a simple model at short-circuit condition:

$$E = \frac{P}{\varepsilon S h} \quad (1)$$

where $\varepsilon$ is the dielectric constant, $S$ and $h$ are the surface area and the thickness of the heterostructures, respectively. Due to the internal electric field, the asymmetric scattering of photoexcited electrons will be expected, and the barrier ($\Delta\Phi$) between HTL and ETL (see in Figure 1(b)) can be furtherly estimated:

$$\Delta\Phi = E \cdot h = \frac{P}{\varepsilon S} \quad (2)$$

Absorbing light excites electron/hole pairs and the internal electric field naturally separate electrons and holes to ETL and HTL (see in Figure 1(c)) to avoid recombination. Comparing to the conventional semiconductor p-n junction with $V_{oc}$ limited by the bandgap of the semiconductor, the $V_{oc}$ of this FE-PV heterostructure will depends on the internal electric field induced by the spontaneous electric polarization of ferroelectric fullerene**.** Thus, ferroelectric fullerenes molecules should have large electric dipole moment to accelerate the separation for getting high photovoltaic energy conversion efficiency.

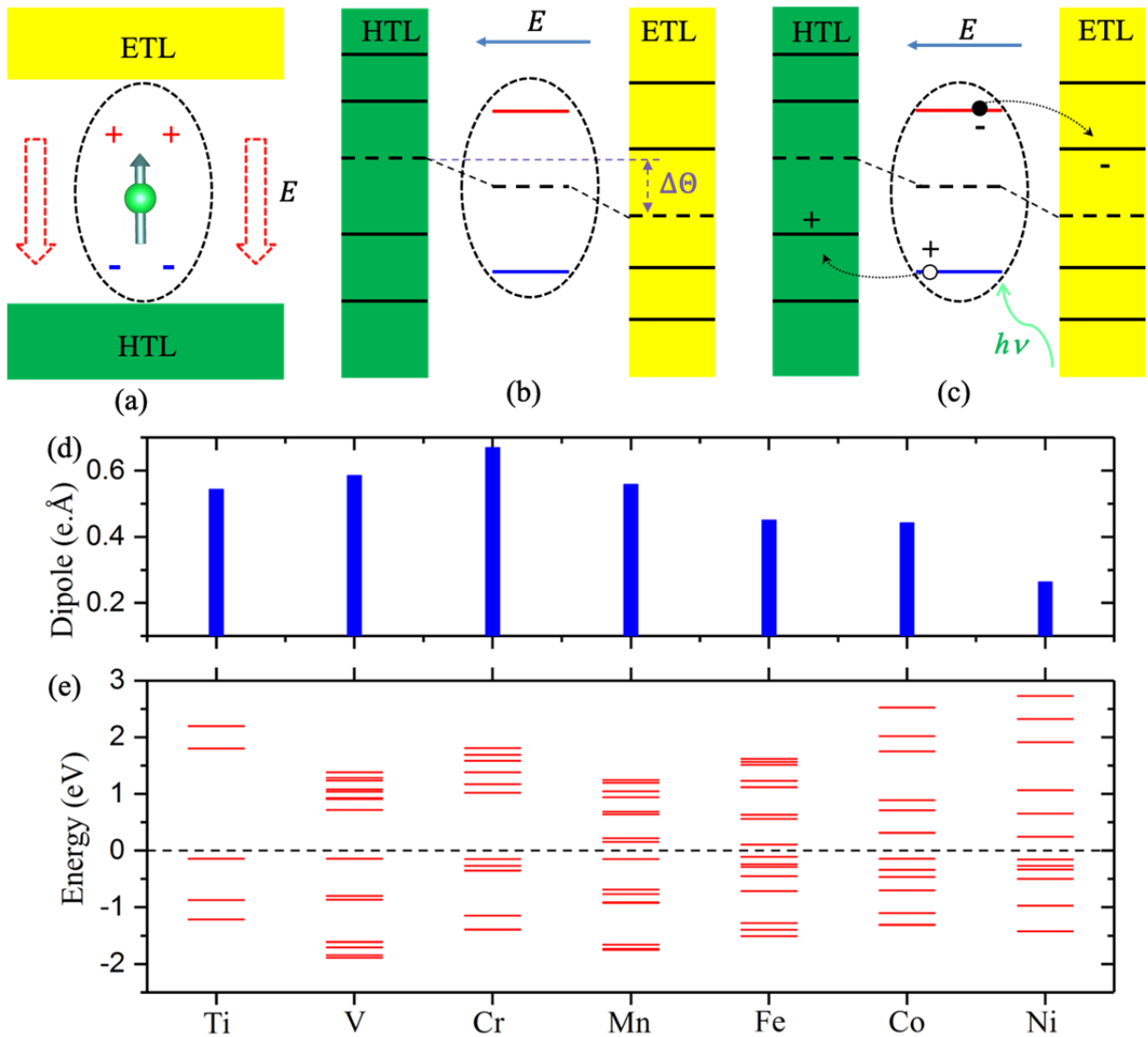

Figure 1. (a) The schematic structure of the ferroelectric and photovoltaic van der Waals heterostructures, in which the endohedral fullerenes were sandwiched between graphene, MoS$_2$ or WS$_2$. (b) The schematic asymmetric barrier induced by the spontaneous electric polarization of ferroelectric fullerene for the heterostructures. (c) The schematic process of optical excitons. (d) (e) The calculated electric dipole moment and energy spectrum of TM@C$_{28}$ (TM= Ti-Ni), respectively.

Here, we demonstrate the performance of such a heterostructure using the smallest stable endohedral fullerenes molecules, TM@$C_{28}$ (TM= Ti-Ni). As shown before,[28-29] most of TM@$C_{28}$ molecules have structural bistability with the $C_3$ symmetry. Using Cr@$C_{28}$ as an example, Cr atom displaces by 0.86 Å as the phase changes (see in Figure S2 in the supplementary materials). From climbing image nudged elastic band (CINEB) calculations with 9 intermediate images between two phases, there is an energy barrier of ~210 meV for the low-energy structural phases of Cr@$C_{28}$ in the conformational transitions. Meanwhile, the separation of positive and negative charge centers (see in the inset of Figure 2(a)) gives rise to electric dipole moments. For example, there is ~1.29 electrons transfer from the Cr atom to the carbon shell for Cr@$C_{28}$ in the low-energy structural phases by using Bader charge analysis, which offers a large dipole moment (~0.67eÅ). Through systematic DFT calculations, the dipole moments and energy spectrum of TM@$C_{28}$ in their low-energy structural phases are obtained. In Figure 1(d) and (e), one may see that Cr@$C_{28}$ has the largest dipole moments and nearly ideal gap (~1.2 eV). Since the GGA approach usually underestimates the band gap, GGA+U (0-4eV) and HSE06 approach were adopted to obtain accurate electronic structures as shown in Figure S1 in the Supplementary Information. One can see that the gap will tends to 1.2eV after the Hubbard U larger than 2eV, even still smaller than the HSE06 calculations (~1.7eV). Considering the balance between calculation costs and accuracy, we will adopt GGA+U (U=2eV) approach in this work. From the projected density of states (PDOS) in Figure 2(a), one can see that a strong hybridization present among Cr and $C_{28}$ orbitals. The highest occupied molecular orbital (HOMO) originates almost completely from the $p$ orbital of Carbon atoms, while the mixture of $p$ orbitals of Carbon atoms and $d_{xy/(x^2-y^2)}$ orbitals of Cr atom form the lowest unoccupied molecular orbital (LUMO). After considering the optical selection rules for electric dipole transitions, we find a prominent optical absorption peak at 1.2 eV that originates as shown in Figure 2(b). Therefore, Cr@$C_{28}$ is a good candidate for both light absorption and carrier separation.

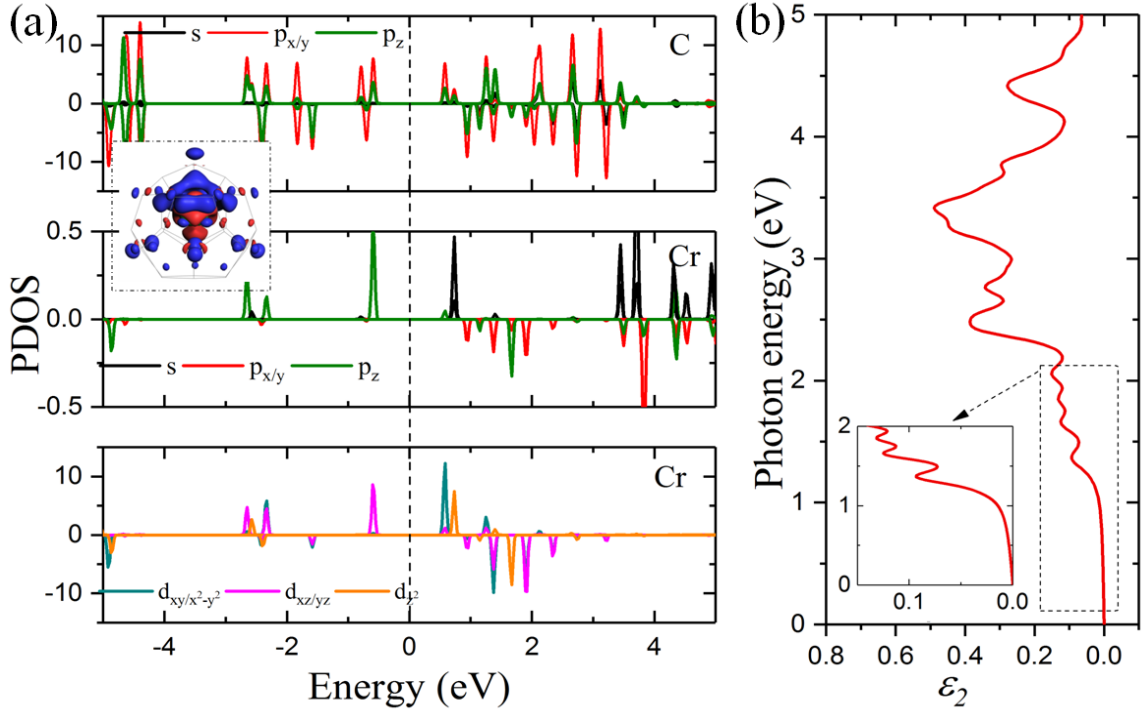

Figure 2. (a) The projected density of states of Cr@$C_{28}$. The inset is its charge redistribution ($\rho_{Cr@C_{28}} - \rho_{C_{28}} - \rho_{Cr}$, red and blue colors indicate charge depletion and accumulation, the isosurface is $2.0\times10^{-3}$e/Å$^3$). (b) The calculated imaginary part of the dielectric function of Cr@$C_{28}$.

In this work, we use graphene as the overlayers (both ETL and HTL) since its remarkable transport and optical properties. As discussed in the previous works,[28] endohedral TM@$C_{28}$ fullerenes prefer the close-packed structure. Thus, a supercell with two 4×4 graphene and one Cr@$C_{28}$ (G/Cr@$C_{28}$/G) is used to simulate the heterostructure and the density of Cr@$C_{28}$ is ~$1.2\times10^{16}$ m$^{-2}$. After being sandwiched between two graphene layers, electrons and holes of

Cr@C$_{28}$ shift toward the top graphene layer (TG) and bottom graphene layer (BG), respectively, as shown in the charge density redistribution in Figure 3(a) and (b). Quantitatively, the Bader charge is ~19.1 mC m$^{-2}$ in graphene and the electric dipole **P** is ~0.19 eÅ. Therefore, a large effective internal electric field is established in the heterostructure and the electrostatic potential difference between TG and BG is ~ 0.43 eV, as shown in the planar-average electrostatic potential in Figure S3 in the supplementary materials. According to Eq. 2, the effective electric field is about 0.04V/Å, which is sufficient to dissociate all charge-transfer excitons and suppress their non-radiative recombination. Meanwhile, the thickness of the heterostructures is ~1.07nm so that carriers can easily transfer from endohedral fullerenes to graphene. To explore the optical absorption properties of the combined system, the dielectric functions $(\varepsilon(\omega) = \varepsilon_1(\omega) + i\varepsilon_2(\omega))$ is calculated and the results are given shown in Figure 3(d). The absorption coefficient $(\alpha(\omega))$ is determined by:[41-42]

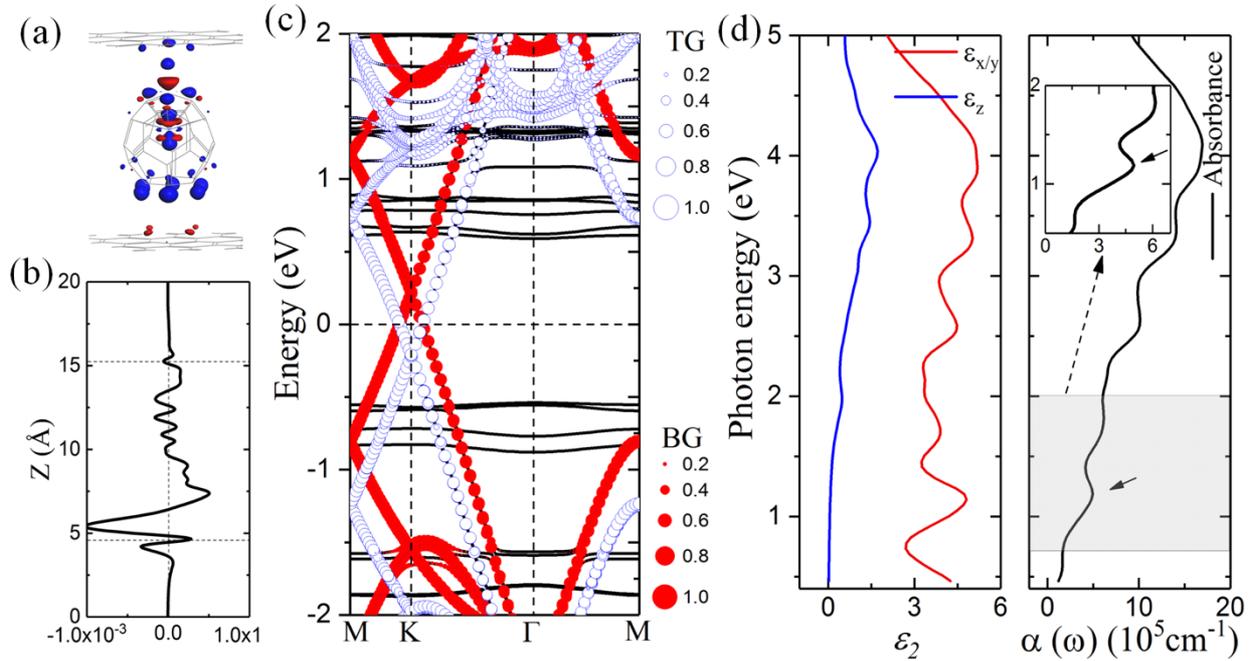

Figure 3. (a) The charge density redistribution of G/Cr@C$_{28}$/G, (Red and blue colors indicate charge depletion and accumulation, the isosurface is $1.0\times10^{-4}$ e/Å$^3$). (b) The corresponding Planar-Average of the charge density redistribution of G/Cr@C$_{28}$/G along z axis. (c) The band structure of G/Cr@C$_{28}$/G. (d) The calculated imaginary part of the dielectric function and absorption coefficient of G/Cr@C$_{28}$/G.

$$\alpha(\omega) = \frac{\sqrt{2}\omega}{c}\left[\sqrt{\varepsilon_1^2 + \varepsilon_2^2} - \varepsilon_1\right]^{\frac{1}{2}}$$

where $\varepsilon_1$ and $\varepsilon_2$ are the real and imaginary parts of the in-plane dielectric function, $\omega$ is the photon frequency. One can see that the optical absorption peak at 1.2eV mostly from Cr@$C_{28}$ still exists since coupling between Cr@$C_{28}$ and graphene is the weak vdW type. It worth to note that the excellent absorption coefficient in the visible light and near-infrared regions can reach the order of ~$10^6$ cm$^{-1}$, which is comparable to other organic perovskite solar cells and two-dimensional semiconductors.[43-46]

We further explore the possibility of using semiconductor transition metal dichalcogenides (MoS$_2$ and WS$_2$) as the overlayers (ETL and HTL). Here, we design two models with either identical overlayers (e.g., MoS$_2$/Cr@$C_{28}$/MoS$_2$) and different overlayers (e.g., WS$_2$/Cr@$C_{28}$/MoS$_2$). A supercell with 3×3 MoS$_2$ (WS$_2$) and one Cr@$C_{28}$ is used to simulate them and the density of Cr@$C_{28}$ in these close-packed heterostructures is ~1.3×$10^{16}$ m$^{-2}$. Their band structures are shown in Figure 4(a) and (d), where one can see larger $\Delta\Phi$ between HTL and ETL. The electrostatic potential difference between two overlayers in MoS$_2$/Cr@$C_{28}$/MoS$_2$ (WS$_2$/Cr@$C_{28}$/MoS$_2$) is ~ 0.69eV (0.95eV) as shown in Figure S3 in the supplementary materials. This suggests a larger internal effective electric field (0.055V/Å, 0.076V/Å) and stronger separation ability for charge-transfer excitons than in G/Cr@$C_{28}$/G. As shown in the Figure 4(c) and (f), the primary optical absorption peak of Cr@$C_{28}$ still present, even with a blueshift of 0.1 eV and 0.3 eV for MoS$_2$/Cr@$C_{28}$/MoS$_2$ and WS$_2$/Cr@$C_{28}$/MoS$_2$. A prominent optical absorption peak at ~2.7eV originating from MoS$_2$ and WS$_2$ are also obvious. Thus, higher photovoltaic energy conversion efficiency can be expected in these two photovoltaic heterostructures, even though the conduction needs to be assisted by electrodes or additional cover layers.

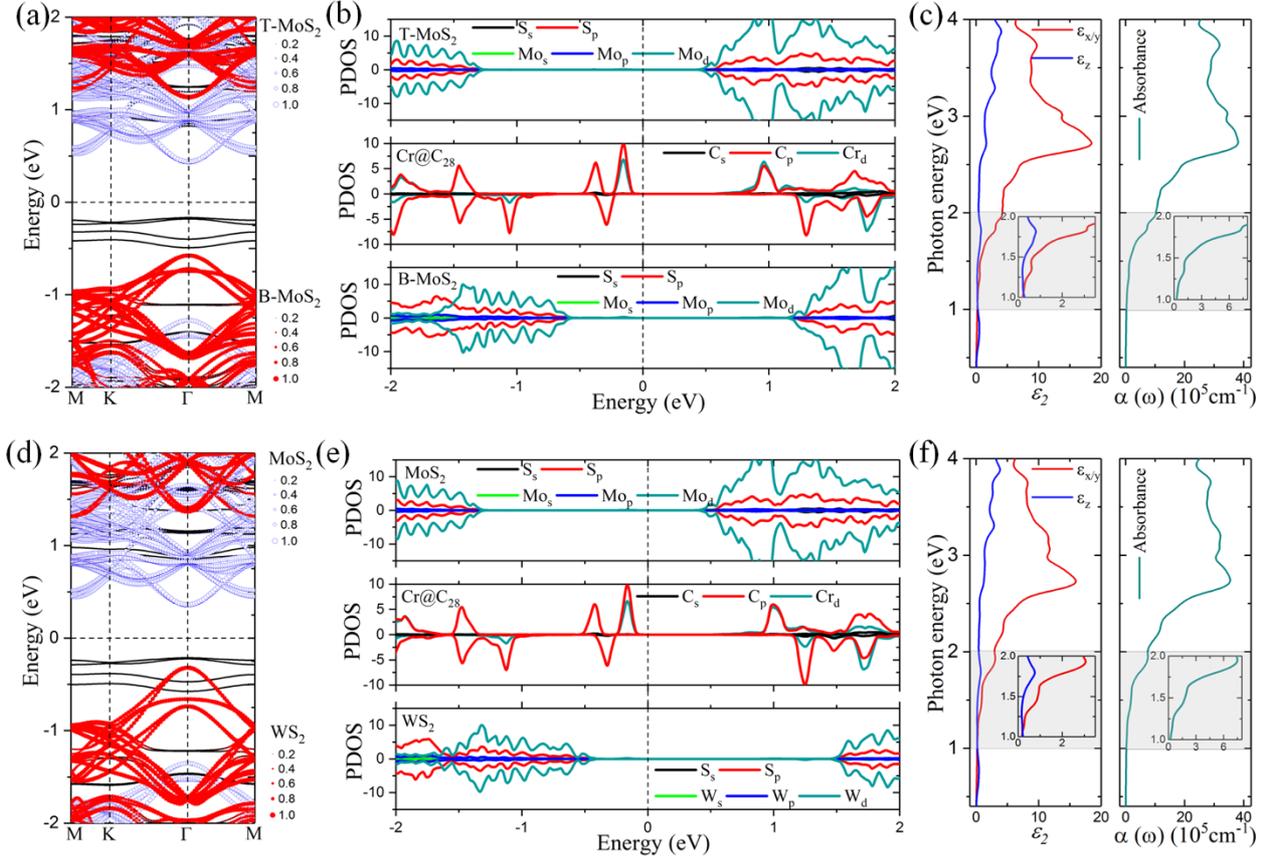

Figure 4. (a)-(c) The band structure, PDOS, calculated imaginary part of the dielectric function and absorption coefficient of MoS$_2$/Cr@C$_{28}$/MoS$_2$, respectively. (d)-(f) The band structure, PDOS, calculated imaginary part of the dielectric function and absorption coefficient of WS$_2$/Cr@C$_{28}$/MoS$_2$, respectively.

In summary, we proposed several vdW heterostructures based on endohedral fullerenes, graphene and transition metal dichalcogenides for photovoltaic applications. Systematic *ab initio* calculations show that they have controllable ferroelectricity and electronic structures by adjusting different transition metal cores. Cr@$C_{28}$ has both large electric dipole moment and idea bandgap (~1.2eV) and hence is a good candidate for ferroelectric-photovoltaic devices. Heterostructures consist of Cr@$C_{28}$, graphene or transition metal dichalcogenides were found to have excellent photovoltaic properties. Due to the large single-molecule quasi-ferroelectricity of Cr@$C_{28}$, the internal electric field in these vdW heterostructures is expected to dissociate all the charge-transfer excitons even considering other factors, such as charge screening, defects and uneven distribution of the endohedral fullerenes. The multilayer photovoltaic heterostructure designs shown in Figure S4 may further enhance the photovoltaic energy conversion efficiency in practical applications. Our findings provide useful insights for the designs of ferroelectric and photovoltaic vdW materials.

This work was supported by US DOE, Basic Energy Science (Grant No. DE-SC0019448). Calculations were performed on parallel computers at NERSC.

**DATA AVAILABILITY**

The data that support the findings of this study are available within the article and its supplementary material.

[1]A. Fujishima, and K. Honda, Nature, **238**, 37 (1972).

[2]A. Goetzberger, and C. Hebling, Sol. Energy Mater. Sol. Cells, **62**, 1 (2000).

[3]C. E. Kennedy, *Review of Mid- to High-Temperature Solar Selective Absorber Materials*, NREL/TP-520-31267, Golden, CO: National Renewable Laboratory (2002).

[4]J. Low, J. Yu, M. Jaroniec, S. Wageh, and A. A. Al-Ghamdi, Adv. Mater. **29**, 1601694 (2017).

[5]J. Gong, C. Li, and M. R. Wasielewski, Chem. Soc. Rev., **48**, 1862 (2019).

[6]L. Wang, L. Huang, W.C. Tan, X. Feng, L. Chen, X. Huang, and K.W. Ang, Small Methods, **2**, 1700294 (2018).

[7]M. A. Green, Physica E, **14**, 11 (2002).

[8]A. Goetzberger, J. Knobloch, B. Voss, and R. Waddington, Crystalline silicon solar cells. Chichester, U.K.: Wiley, 1998.

[9]C. Battaglia, A. Cuevas, and S. De Wolf, Energy Environ. Sci., **9**, 1552 (2016).

[10]D. E. Carlson, and C. R. Wronski, Appl. Phys. Lett. **28**, 671 (1976).

[11]C. Zhang, Y. Song, M. Wang, M. Yin, X. Zhu, L. Tian, H. Wang, X. Chen, Z. Fan, L. Lu, and D. Li, Adv. Funct. Mater. **27**, 1604720 (2017).

[12]N. J. Jeon, J. H. Noh, W. S. Yang, Y. C. Kim, S. Ryu, J. Seo, and S. Seok, Nature **517**, 451 (2014).

[13]W. J. Yin, T. Shi, and Y. Yan, Appl. Phys. Lett. **104,** 063903 (2014).

[14]H. Ren, S. Yu, L. Chao, Y. Xia, Y. Sun, et al., Nat. Photonics **14**, 154 (2020).

[15]C. H. M. Chuang, P. R. Brown, V. Bulović, and M. G. Bawendi, Nat. Mater. **13**, 796 (2014).

[16]A. Polman, M. Knight, E. C. Garnett, B. Ehrler, and W. C. Sinke, Science **352**, 6283 (2016).

[17]L. Pintilie, I. Vrejoiu, G. Le Rhun, and M. Alexe, J. Appl. Phys., 101, 064109 (2007).

[18]Y. Yuan, Z. Xiao, B.Yang, and J. Huang, J. Mater. Chem. A, **2**, 6027 (2014).

[19]B. Kang, B. K. Rhee, G.-T. Joo, S. Lee, and K.-S. Lim, Opt. Commun., 266, 203 (2006).

[20]M. Ichiki, R. Maeda, Y. Morikawa, Y. Mabune, T. Nakada, and K. Nonaka, Appl. Phys. Lett., 84,

395 (2004).

[21]T. Choi, S. Lee, Y. J. Choi, V. Kiryukhin, and S. W. Cheong, Science, 324, 63 (2009).

[22]M. Qin, K. Ao, and Y. C. Liang, Appl. Phys. Lett. **93**, 122904 (2008).

[23]I. Grinberg, D. West, M. Torres, G. Gou, D. M. Stein, et al., Nature **503**, 509 (2013).

[24]D. S. Bethune, R. D. Johnson, J. R. Salem, M. S. de Vries, and C. S. Yannoni, Nature **366**, 123 (1993).

[25]J. Cioslowski, and A. Nanayakkara, Phys. Rev. Lett. **69**, 2871 (1992).

[26]P. W. Dun, N. K. Kaiser, M. Mulet-Gas, A. Rodríguez-Fortea, J. M. Poblet, et al., J. Am. Chem. Soc. **134**, 9380 (2012).

[27]K. Zhang, C.Wang, M. Zhang, Z. Bai, F. F. Xie, et al., Nat. Nanotechnol. **15**, 1019 (2020).

[28]J. Li, and R. Q. Wu, Nanoscale, **13**, 12513 (2021).

[29]J. Li, L. Gu, and R. Q. Wu,  arXiv:2012.06141.

[30]L. Yang, J. Deslippe, C. H. Park, M. L. Cohen, and S. G. Louie, Phys. Rev. Lett. **103**, 186802 (2009).

[31]S. K. Behura, C. Wang, Y. Wen, and V. Berry, Nat. Photonics **13**, 312 (2019).

[32]C. M. Went, J. Wong, P. R. Jahelka, M. Kelzenberg, S. Biswas, et al., Sci. Adv. 5, eaax6061 (2019).

[33]G. Li, H. T. Zhou, L. D. Pan, Y. Zhang, J. H. Mao, et al., Appl. Phys. Lett. **100**, 013304 (2012).

[34]K. Kim, T. H. Lee, E. J. G. Santos, P. S. Jo, A. Salleo, et al., ACS Nano, **9**, 5922 (2015).

[35]R. Mirzayev, K. Mustonen, M. R. A. Monazam, A. Mittelberger, Sci. Adv. **3**, e1700176 (2017).

[36]J. P. Perdew, K. Burke, and M. Ernzerhof, Phys. Rev. Lett. **77**, 3865 (1996).

[37]P. E. Blochl, Phys. Rev. B **50**, 17953 (1994).

[38]G. Kresse, and D. Joubert, Phys. Rev. B **59**, 1758 (1999).

[39]S. L. Dudarev, G. A. Botton, S. Y. Savrasov, C. J. Humphreys, A. P. Sutton, Phys. Rev. B **57**, 1505

(1998).

[40]S. Grimme, J. Antony, S. Ehrlich, and S. Krieg, J. Chem. Phys. **132**, 154104 (2010).

[41]S. Saha, T. Sinha, and A. Mookerjee, Phys. Lett. B, **62**, 8828 (2000).

[42]X. Zhang, X. Zhao, D. Wu, Y. Jing, and Z. Zhou, Adv. Sci., 3, 1600062 (2016).

[43]N. J. Jeon, J. H. Noh, Y. C. Kim, W. S. Yang, S. Ryu, et al., Nat. Mater., **13**, 897 (2014).

[44]M. Shirayama, H. Kadowaki, T. Miyadera, T. Sugita, M. Tamakoshi, et al., Mater. Sci., 5, 014012 (2016).

[45]Z. Xiao, W. Meng, J. Wang, D. B. Mitzi, Y. Yan, Mater. Horiz. 4, 206 (2017).

[46]S. Sun, F. Meng, H. Wang, H. Wang, and Y. Ni, J. Mater. Chem. A, 6, 11890 (2018).